\documentclass[%
 aip,
 amsmath,amssymb,
reprint,%
]{revtex4-1}

\usepackage{graphicx}
\usepackage{dcolumn}
\usepackage{subcaption}
\usepackage{comment}
\usepackage{caption}
\usepackage{hyperref}
\hypersetup{
    colorlinks=true,
    linkcolor=blue,
    citecolor =blue,
    urlcolor=blue,
    }
\usepackage[mathlines]{lineno}

\usepackage[utf8]{inputenc}
\usepackage[T1]{fontenc}
\usepackage{txfonts}
\usepackage{etoolbox}

\usepackage{verbatim}

\makeatletter
\def\@email#1#2{%
 \endgroup
 \patchcmd{\titleblock@produce}
  {\frontmatter@RRAPformat}
  {\frontmatter@RRAPformat{\produce@RRAP{*#1\href{mailto:#2}{#2}}}\frontmatter@RRAPformat}
  {}{}
}%
\makeatother

\begin{document}

\preprint{AIP/123-QED}
\title{Amplitude Noise Cancellation of Microwave Tones}

\author{Joe Depellette}
\affiliation{QTF Centre of Excellence, Department of Applied Physics, Aalto University, FI-00076 Aalto, Finland}
\author{Ewa Rej}
\affiliation{QTF Centre of Excellence, Department of Applied Physics, Aalto University, FI-00076 Aalto, Finland}
\author{Matthew Herbst}
\affiliation{QTF Centre of Excellence, Department of Applied Physics, Aalto University, FI-00076 Aalto, Finland}
\author{Richa Cutting}
\affiliation{QTF Centre of Excellence, Department of Applied Physics, Aalto University, FI-00076 Aalto, Finland}
\author{Yulong Liu}
\affiliation{Beijing Academy of Quantum Information Sciences, Beijing 100193, China}
\author{Mika A. Sillanpää}
\affiliation{QTF Centre of Excellence, Department of Applied Physics, Aalto University, FI-00076 Aalto, Finland}

\email{mika.sillanpaa@aalto.fi}

\date{2 June 2025}

\begin{abstract}
Carrier noise in coherent tones limits sensitivity and causes heating in many experimental systems, such as force sensors, time-keeping, and studies of macroscopic quantum phenomena. Much progress has been made to reduce carrier noise using phase noise cancellation techniques, however, in systems where amplitude noise dominates, these methods are ineffective. Here, we present a technique to reduce amplitude noise from microwave generators using feedback cancellation. The method uses a field-programmable gate array (FPGA) to reproduce noise with a tunable gain and time delay, resulting in destructive interference when combined with the original tone. The FPGA additionally allows for tuning of the frequency offset and bandwidth in which the noise is canceled. By employing the cancellation we observe 13\,dB of noise power reduction at a 2\,MHz offset from a 4\,GHz microwave tone, lowering the total noise to the phase noise level. To verify its applicability we utilize the setup in a microwave optomechanics experiment to investigate the effect of generator noise on the sideband cooling of a 0.5\,mm silicon nitride membrane resonator. We observe that with our technique the rate of externally induced cavity heating is reduced by a factor of 3.5 and the minimum oscillator occupation is lowered by a factor of 2. This method broadens the field of noise cancellation techniques, where amplitude noise is becoming an increasingly important consideration in microwave systems as phase noise performances improve over time.
\end{abstract}

\maketitle

\section{Introduction}

Low noise microwave sources and lasers are essential for a wide range of technological applications and studies of fundamental physics, particularly in the field of cavity optomechanics, which has enabled feedback control of mechanical motion \cite{cohadon1999cooling,kleckner2006sub,poggio2007feedback,wilson2015measurement,rossi2018measurement,wang2023fast, rej2025near}, motional entanglement between mechanical resonators \cite{jost2009entangled,ockeloen2018stabilized, riedinger2018remote}, and studying the relationship between quantum mechanics and gravity \cite{belenchia2016testing,al2018optomechanical,liu2021gravitational,westphal2021measurement,pedernales2022enhancing}. These all require the use of coherent tones to manipulate mechanical resonators.

Inevitably, all sources produce noise at offsets from their tone's frequency. Generator noise, or laser noise in the case of optical experiments, causes heating of systems which use sideband cooling, often limiting the minimum achievable occupation number of the oscillator \cite{schliesser2008sbcooling, Rabl2009phase-noise, kippenberg2013phase, safavi2013laser, Meyer2019levitated}. This problem is typically addressed by direct filtering \cite{teufel2011sideband, liu2022quantum}, where band-stop filters are used to reduce noise at the frequency of interest, often at an offset of one mechanical frequency from the carrier. However, this technique encounters a problem when the mechanical frequency becomes comparable to the filter bandwidth; one cannot filter so close to the central tone without also reducing the tone’s power. An obvious solution would be to decrease the filter bandwidth, but for mechanical frequencies in the low-kilohertz regime this approach requires tunable cavities cooled to cryogenic temperatures \cite{c2016piezoelectric,souris2017tuning,clark2018cryogenic,potts2020strong}, which are yet to achieve such high quality factors.

An alternative to directly filtering noise is to implement cancellation techniques. Most often, the dominant source of generator noise is phase noise, caused by short-term fluctuations in frequency, and previous work has focused on cancellation in this regime \cite{hati2008cancellation, aflatouni2010design, kenig2012passive, gharavi2013new, chen2015high, thijssen2017feedforward, parniak2021high}. However, there are certain cases where amplitude noise, characterized by short-term fluctuations in output power, dominates over the phase noise. This regime is often overlooked, with little work being done to specifically target amplitude noise. A notable exception is the use of cryogenic resonators to reduce both phase and amplitude fluctuations at several kilohertz offsets from a microwave tone \cite{ivanov2021noise}.

Here, we present a technique based on manipulating generator noise with a field-programmable gate array (FPGA), and cancel in a regime where amplitude noise dominates. We demonstrate sizable reductions in noise power, along with substantial control over the central frequency and bandwidth of cancellation. Additionally, the setup is used in a standard sideband cooling experiment, involving a micro-mechanical resonator coupled to a 3D cavity within a dilution refrigerator. The results are compared to those obtained using direct noise filtering.

The additional freedoms introduced by our technique when compared to cavity filters, such as tunable bandwidth, may enable future experiments that would otherwise be more challenging. One example is cancellation at sub-kilohertz offsets from carrier tones when combined with phase noise cancellation, which is useful for studies involving low frequency mechanical resonators. Additionally, directly targeting amplitude noise has applications in sensitive phase noise measurements, where amplitude noise can mask low phase noise levels.

\section{Noise Cancellation}

\begin{figure}
    \centering
    \includegraphics[width=0.9\linewidth]{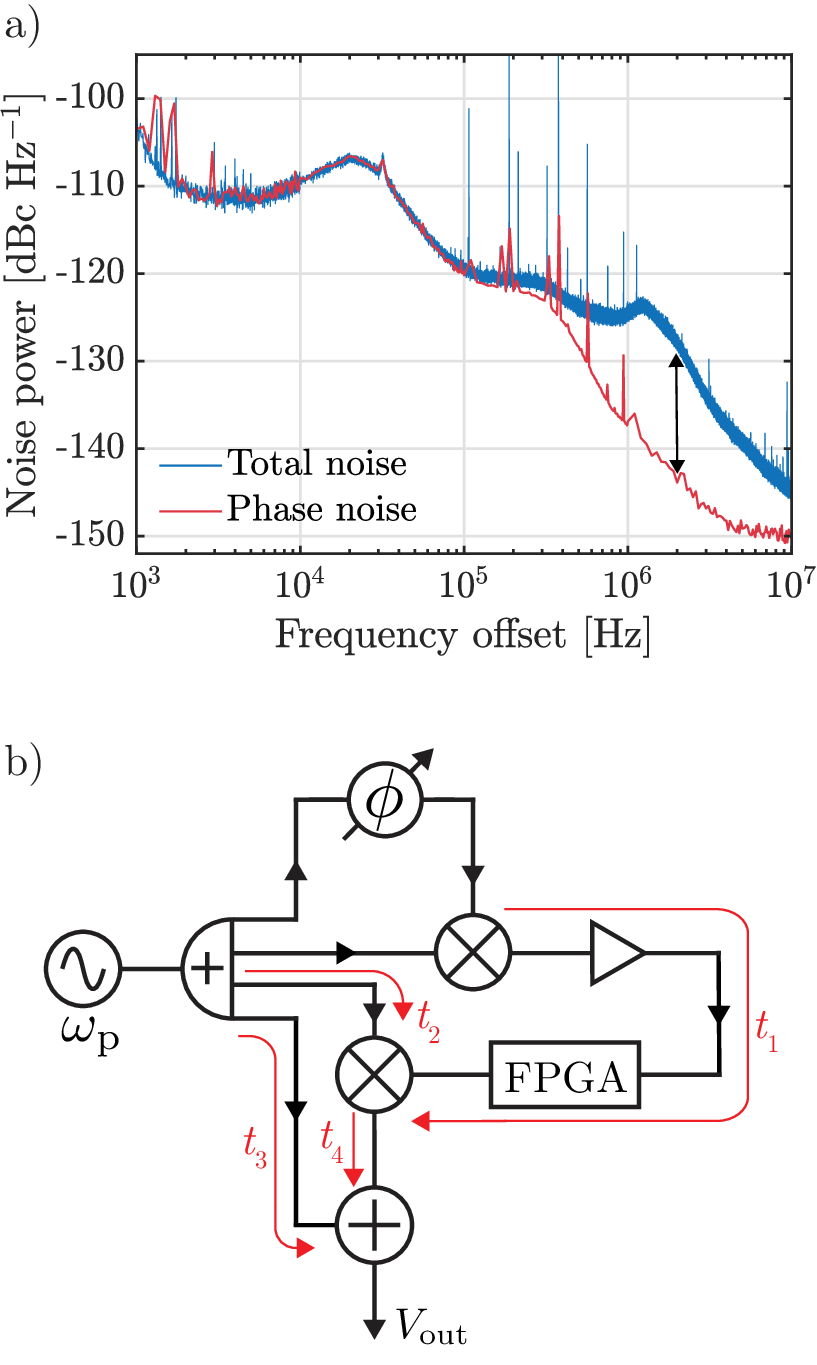}
    \caption{(a) Comparison of the measured total noise and phase noise of a microwave tone: At low offsets the phase noise is the dominant contributor to the total noise. Cancellation is done at an offset of 2\,MHz, where the difference between the two noise powers is significant. (b) Schematic of the circuit used for amplitude noise cancellation.}
    \label{fig:CancelSetup}
\end{figure}

We now turn to our experimental setup, where we use a standard commercial signal generator to produce tones in the range 4\,GHz to 5\,GHz. Using signal analyzer measurements with a phase noise measurement function, we demonstrate that for frequency offsets from approximately 200\,kHz to 10\,MHz, the total noise from the generator is larger than the phase noise, shown in Fig.~\ref{fig:CancelSetup}(a). This excess noise is attributed to amplitude noise present in the region, and the cancellation technique to address it is described next.

\subsection{Cancellation circuit}

We start by modeling the microwave tone as a sinusoidal voltage with a modulated amplitude

\begin{equation}
    V_{\mathrm{RF}}(t) = [V_0 + V(t)]\sin(\omega_0 t)\,.
\end{equation}

\noindent The carrier tone has a single frequency $\omega_0$ and amplitude $V_0 + V(t)$, which includes an average voltage $V_0$ and a time dependent component $V(t)$ describing the noise. We work with the assumption that $|V_0| \gg |V(t)|$. We approximate the amplitude noise as a summation of sinusoidal terms of the form

\begin{equation}
    V(t) = \sum_i \hat{V}_i\sin(\omega_i t + \phi_i)\,,
\end{equation}

\noindent where $\omega_i$ are the frequency offsets from the carrier and each term has a frequency dependent magnitude $\hat{V}_i$ and phase $\phi_i$.

In our setup, the tone is split into four paths of equal power, two of which are sent to a mixer, see Fig.~\ref{fig:CancelSetup}(b). Homodyne mixing of these tones produces DC and high frequency components, while also producing a term proportional to $V(t)$, which describes amplitude noise at all frequencies. An FPGA is used to apply a band-pass filter, removing the DC and high frequency components, leaving

\begin{equation}    \label{Eq:VIF}
    V_\mathrm{IF}(t) = G_1V_0\Tilde{V}(t)\,,
\end{equation}

\noindent where $G_1$ includes the mixer conversion gain in addition to amplification before the FPGA. Here, $\Tilde{V}$ is the noise after filtering, which picks up a frequency dependent magnitude and phase based on the shape of the filter. Describing the filter as a complex function $H(\omega)$ allows the noise to be expressed as

\begin{equation}
    \Tilde{V}(t) = \sum_i |H(\omega_i)|\hat{V}_i\sin\left[\omega_i t + \phi_i + \arg\{ H(\omega_i)\}\right]\,.
\end{equation}

\noindent Filtering of the input voltage here is necessary to avoid saturation of the FPGA due to high powers. The noise is amplified and given a time delay by the FPGA before being mixed with the original tone. For cancellation the optimum FPGA gain is given by

\begin{equation}    \label{eq:G_RP}
    G = -\frac{1}{G_\mathrm{tot}V_0^2}\,,
\end{equation}

\noindent where $G_\mathrm{tot}$ is the sum of all other gains and attenuation the noise experiences in the process. The noise and the third carrier tone path arrive at a second mixer, which outputs

\begin{equation}
    V_\mathrm{up}(t) = -\Tilde{V}(t - t_1)\sin[\omega_0 (t - t_2)]\,,
\end{equation}

\noindent when the gain given by Eq.~(\ref{eq:G_RP}) is used. Here, we take into account the time delays experienced by the noise ($t_1$) and the carrier ($t_2$), where the delay $t_1$ is tunable via the FPGA. The mixed-up noise experiences a delay $t_4$ before being combined with the final path of the carrier tone (which arrives with a delay $t_3$) and destructively interferes with the original amplitude noise, producing

\begin{multline}    \label{eq:Vout}
    V_\mathrm{out}(t) = V_0\sin[\omega_0 (t - t_3)] + V(t - t_3)\sin[\omega_0 (t - t_3)]\\
    - \Tilde{V}(t - t_1')\sin[\omega_0 (t - t_2')]\,,
\end{multline}

\noindent where $t_{1,2}' = t_{1,2} + t_4$. We now consider the interference between terms 2 and 3 in Eq.~(\ref{eq:Vout}) at a single frequency, $(\omega_0 + \omega_1)$, where the resulting component is proportional to

\begin{multline}
    |H(\omega_1)|\cos[(\omega_0 + \omega_1)t - \omega_0 t_2' - \omega_1 t_1' + \arg\{H(\omega_1)\}]\\
    - \cos[(\omega_0 + \omega_1)(t - t_3)]\\
    = A(\omega_1)\cos[(\omega_0 + \omega_1)t + \theta(\omega_1)]\,,
\end{multline}

\noindent where $A(\omega_1)$ and $\theta(\omega_1)$ are the resulting magnitude and phase after interference. In our experiment, the carrier tone exhibits both phase and amplitude noise, such that the total noise power at offset $\omega_1$ is

\begin{equation}
    S_\mathrm{tot}(\omega_1) = S_\mathrm{\phi}(\omega_1) + S_\mathrm{amp}(\omega_1)\,,
\end{equation}

\noindent where $S_\mathrm{\phi}(\omega_1)$ and $S_\mathrm{amp}(\omega_1)$ are the phase and amplitude noise powers respectively. At the output of the cancellation circuit, the interference modifies the amplitude noise such that the total power becomes

\begin{equation}    \label{Eq:S'}
    S'_\mathrm{tot}(\omega_1) = S_\mathrm{\phi}(\omega_1) + A^2(\omega_1)S_\mathrm{amp}(\omega_1)\,.
\end{equation}

\noindent We now consider the shape of the noise profile in the absence of filtering, such that $H = 1$. The factor by which the noise is modified takes the form

\begin{equation}
    A^2(\omega_1) = 4 \sin^2\left[ \frac{\omega_0}{2}(t_3 - t_2') + \frac{\omega_1}{2}(t_3 - t_1') \right]\,.
\end{equation}

\noindent Around the cancellation frequency $(\omega_0 + \omega_1)$, the width of noise profile, where $A^2(\omega_1) < 1$, is

\begin{equation}
    \Delta\omega = \frac{2\pi}{3\left|t_3 - t_1'\right|}\,,
\end{equation}

\noindent demonstrating that closely matched delays are required for cancellation over a broad frequency range. Based on the analysis above we predict that broadband canceling is possible when all the time delays are matched. This, however, may introduce problems with attenuation of the carrier signal due to the longer cable lengths required.

For the FPGA we use a Red Pitaya STEMlab 125-14 digitizer board and PyRPL software \cite{neuhaus2024python}. Additionally, a manually operated phase shifter before the homodyne mixing ensures that the resulting voltage follows Eq.~(\ref{Eq:VIF}). Here, the delays $t_2$ and $t_3$ are primarily determined by cable lengths, and are on the order of 5\,ns. The FPGA latency sets the lower limit on $t_1$ to be 210\,ns.

Two limitations on the minimum achievable noise power can be considered. Firstly, as in our experiment, any phase noise in the carrier tone sets a lower limit on the canceled total noise. Secondly, the FPGA has an intrinsic noise floor which, in our case, is $-119$\,dBm Hz$^{-1}$. To cancel noise with a lower power than this, it must first be amplified before the FPGA input. Amplification introduces additional noise which, above a certain gain, becomes comparable to the magnitude of $V_{\mathrm{IF}}$ and affects the circuit output. The intrinsic generator noise in our experiment is large enough such that we are not limited by this case.

\begin{figure*}[t!]
    \centering
    \includegraphics[width=0.99\linewidth]{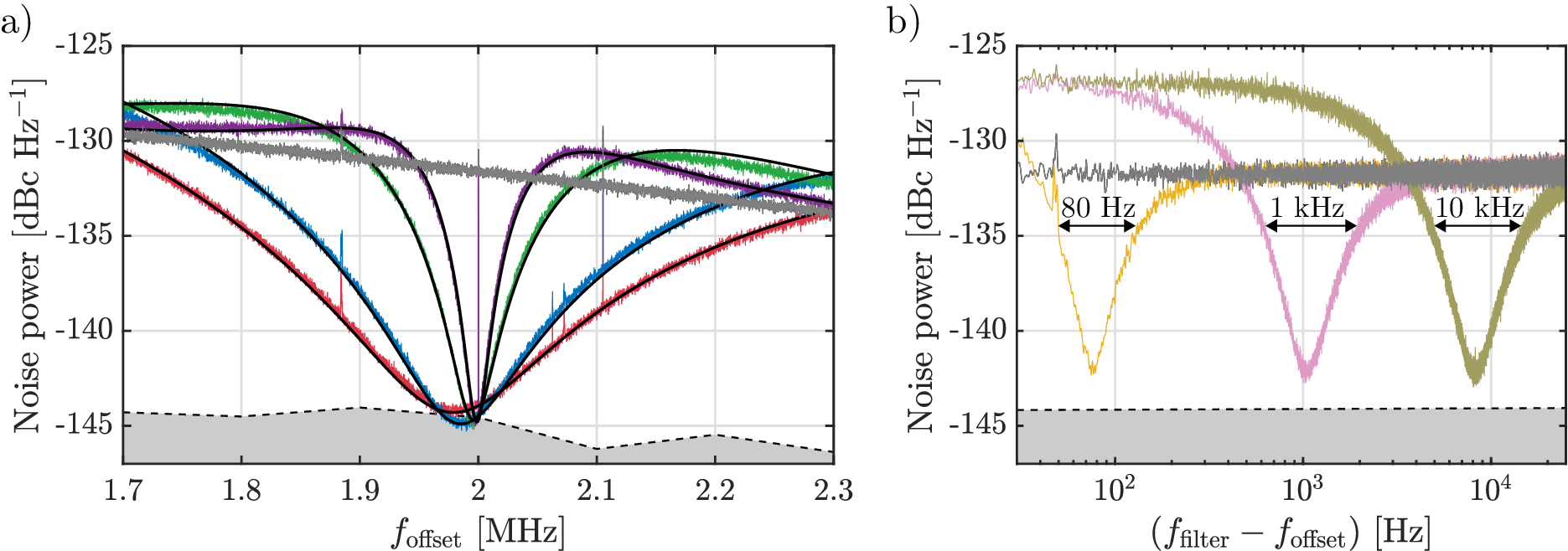}
    \caption{Noise cancellation of a 4\,GHz tone. The dark gray line depicts the uncanceled total noise and the shaded light gray region shows the phase noise: (a) With optimized parameters we measure amplitude noise cancellation reaching the phase noise level for filter bandwidths [0.1; 0.2; 1; 2]\,MHz from narrowest to widest, where $f_\mathrm{offset}$ is the frequency offset from the carrier tone. Black lines are fits to the data based on Eq.~(\ref{Eq:S'}). (b)~ Canceling with smaller filter bandwidths [0.1; 1; 10]\,kHz from left to right. The effect of an unoptimized phase is presented here with asymmetric noise profiles and a shift in the frequency of minimum power to be off-center from the filter. Here, $f_\mathrm{filter}$ is the central frequency of the band-pass filter. The resulting FWHM of each profile are labeled in the plot.}
    \label{fig:cancelling}
\end{figure*}

\subsection{Cancellation results}

We begin our experiment with the aim to characterize the effectiveness of the cancellation technique to reduce noise at a selected frequency offset. A 4\,GHz microwave tone is passed through the cancellation circuit and a signal analyzer is used to measure the noise spectra at the output.

The amplitude noise heavily dominates over the phase noise at a 2\,MHz offset, making this a good target to test the canceling, see Fig.~\ref{fig:CancelSetup}(a). The FPGA parameters were tuned for maximum decrease in total noise and a symmetric profile. The measured output power at a 2\,MHz offset, our chosen FPGA filter frequency, is shown in Fig.~\ref{fig:cancelling}(a) for a selection of filter bandwidths.

In the instance of optimized parameters, we show that the total noise power is reduced by approximately 13\,dB from its uncanceled value at the central frequency of the filter, lowering it to the phase noise level. Additionally, we demonstrate that the width of the canceled region is controlled by the filter bandwidth. Fits to the data show reasonable agreement with the noise profile predicted by Eq.~(\ref{Eq:S'}) when a 4th order filter is used. Here, the delays $t_i$ are used as fitting parameters, which take reasonable values based on the cable lengths and components used in the circuit. The uncanceled noise powers used for the fits are measured data, as shown in Fig.~\ref{fig:cancelling}.

We also present canceling for smaller filter bandwidths, shown in Fig.~\ref{fig:cancelling}(b). In this case we use an unoptimized phase difference between the first two signal paths to demonstrate its effect and to display noise profiles with widths varying over a large frequency range. The profiles are asymmetric and the noise minima appear off-center from the filter frequency. Regardless, we show that the minimum achievable noise power is largely unaffected by the filter bandwidth, demonstrating the feasibility of effective canceling with sub-kilohertz profile widths.

\section{Sideband Cooling}

Next, we demonstrate the cancellation technique in an optomechanics experiment to mitigate heating caused by generator noise. 

\subsection{Optomechanical theory}

Optomechanical systems rely on the coupling between a cavity with resonance frequency $\omega_\mathrm{c}$ and a mechanical oscillator with resonance frequency $\omega_\mathrm{m}$. The energy loss rate of the cavity $\kappa$ features two components: internal losses ($\kappa_\mathrm{i}$) to the environment and external losses ($\kappa_\mathrm{e}$) to the capacitively coupled transmission line, such that $\kappa = \kappa_\mathrm{i} + \kappa_\mathrm{e}$. The intrinsic loss rate of the mechanical oscillator is $\Gamma_\mathrm{m}$. The cavity frequency is parametrically coupled to the position $x$ of the oscillator, and the single-photon optomechanical coupling strength is

\begin{equation}
    g_0 = \frac{\mathrm{d}\omega_\mathrm{c}}{\mathrm{d}x}x_\mathrm{ZP}\,,
\end{equation}

\noindent where $x_\mathrm{ZP}$ is the zero-point motion of the oscillator.  The coupling is enhanced by inputting a tone to the cavity with frequency $\omega_{\mathrm{p}}$ and detuning $\Delta = \omega_{\mathrm{p}} - \omega_{\mathrm{c}}$, populating the cavity with

\begin{equation}
    n_\mathrm{cav} = \frac{\kappa_\mathrm{e}}{\Delta^2 + (\kappa/2)^2}\frac{P}{\hbar \omega_\mathrm{p}}
\end{equation}

\noindent coherent photons, where $P$ is the input power. The quantity $g_0\sqrt{n_{\mathrm{cav}}}$ is then referred to as the effective coupling. The pump facilitates Stokes and anti-Stokes scattering between the injected photons and oscillator phonons. Furthermore, a red-detuned pump ($\Delta < 0)$ favors the anti-Stokes scattering due to the cavity susceptibility, resulting in cooling of the oscillator, known as sideband cooling. This leads to an optical damping, increasing the damping rate of the oscillator by

\begin{multline}    \label{eq:Gammaopt}
    \Gamma_{\mathrm{opt}} = g_0^2n_{\mathrm{cav}}\left( \frac{\kappa}{\kappa^2/4 + (\Delta + \omega_\mathrm{m})^2}\right.\\
    - \left.\frac{\kappa}{\kappa^2/4 + (\Delta - \omega_\mathrm{m})^2} \right)\,,
\end{multline}

\noindent such that the total damping rate is $\Gamma_\mathrm{eff} = \Gamma_\mathrm{m} + \Gamma_\mathrm{opt}$. Here, the oscillator is coupled to two thermal baths, namely the cryostat environment with phonon population $n_\mathrm{m}^T$ and the cavity with thermal photon population $n_\mathrm{c}^T$, see Fig.~\ref{fig:CavSchematic}(a). The cavity itself is coupled to two environments: the internal bath with population $n_\mathrm{i}$, and external bath with population $n_\mathrm{e}$. The internal bath is related to the temperature of the cryostat by Bose-Einstein statistics, while the external bath is related to noise injected from the transmission line, primarily due to generator noise. The contributing weights to the total cavity noise are given by the couplings to each of these baths, resulting in

\begin{equation}    \label{eq:ncT}
    n_\mathrm{c}^T = \frac{\kappa_\mathrm{i}}{\kappa}n_\mathrm{i} +  \frac{\kappa_\mathrm{e}}{\kappa}n_\mathrm{e}\,.
\end{equation}

\noindent When large microwave powers are present inside the cavity, heating of the internal bath can occur, raising the value of $n_\mathrm{i}$ to well above the corresponding cryostat temperature. This is called technical heating, a longstanding observation in optomechanics experiments \cite{regal2008measuring, Schwab2010, pirkkalainen2015squeezing, ockeloen2018stabilized, mercier2021quantum, rej2025near} which lacks a robust theoretical description. Naturally, $n_\mathrm{e}$ also increases with power due to the corresponding increase in generator noise. In addition, the oscillator experiences a fundamental backaction from the cavity, contributing

\begin{equation}
    n_\mathrm{ba} = \left(\frac{(\kappa/2)^2 + (\Delta - \omega_\mathrm{m})^2}{(\kappa/2)^2 + (\Delta + \omega_\mathrm{m})^2} - 1 \right)^{-1}
\end{equation}

\noindent quanta, which manifests from quantum fluctuations in the cavity in addition to thermal cavity noise. The final phonon population of the oscillator during sideband cooling is then given by

\begin{equation}    \label{eq:nm}
    n_\mathrm{m} = \frac{\Gamma_\mathrm{m}}{\Gamma_\mathrm{eff}}n_\mathrm{m}^T + \frac{\Gamma_\mathrm{opt}}{\Gamma_\mathrm{eff}}\left[n_\mathrm{c}^T\left( 1 + 2n_\mathrm{ba}\right) + n_\mathrm{ba}\right]\,.
\end{equation}

\noindent One can observe that by increasing the optical damping, the relative coupling shifts from favoring the hot phonon bath to favoring the comparatively cold cavity bath, thereby decreasing the effective temperature. A limit on the minimum phonon number, which occurs at high optical damping when $\Gamma_\mathrm{eff} \approx \Gamma_\mathrm{opt}$, is determined by the cavity noise and backaction. A useful quantity when describing optomechanical systems is the cooperativity, given by

\begin{equation}
    C = \frac{\Gamma_\mathrm{opt}}{\Gamma_\mathrm{m}}\,.
\end{equation}

When the pump tone is reflected from the cavity, the optomechanical interaction induces sidebands in the output spectrum. The mechanical oscillations are imprinted on the sidebands, and thus the information in the sidebands serves as our characterization for the injected generator noise. It is noteworthy that particularly in the presence of externally injected noise, the sidebands do not exactly reproduce the spectrum of the oscillator itself, since interference between noise and the emitted signal at the device level distorts the spectral profiles.

The theoretical background presented in this section holds for a general optomechanical system in the linear and weak coupling regimes. In the resolved sideband limit $(\omega_\mathrm{m} \gg \kappa)$ with no external heating there are simple analytical formulae for the output spectra \cite{Schwab2010,teufel2011sideband}. Such expressions also exist in the unresolved sideband limit $(\omega_\mathrm{m} \ll \kappa)$. In our intermediate situation where $\omega_\mathrm{m} \approx \kappa$, it is necessary to apply corrections to these two limiting cases, for which we use a fully numerical approach.

\subsection{Experimental setup}

\begin{figure}
    \centering
    \includegraphics[width=0.8\linewidth]{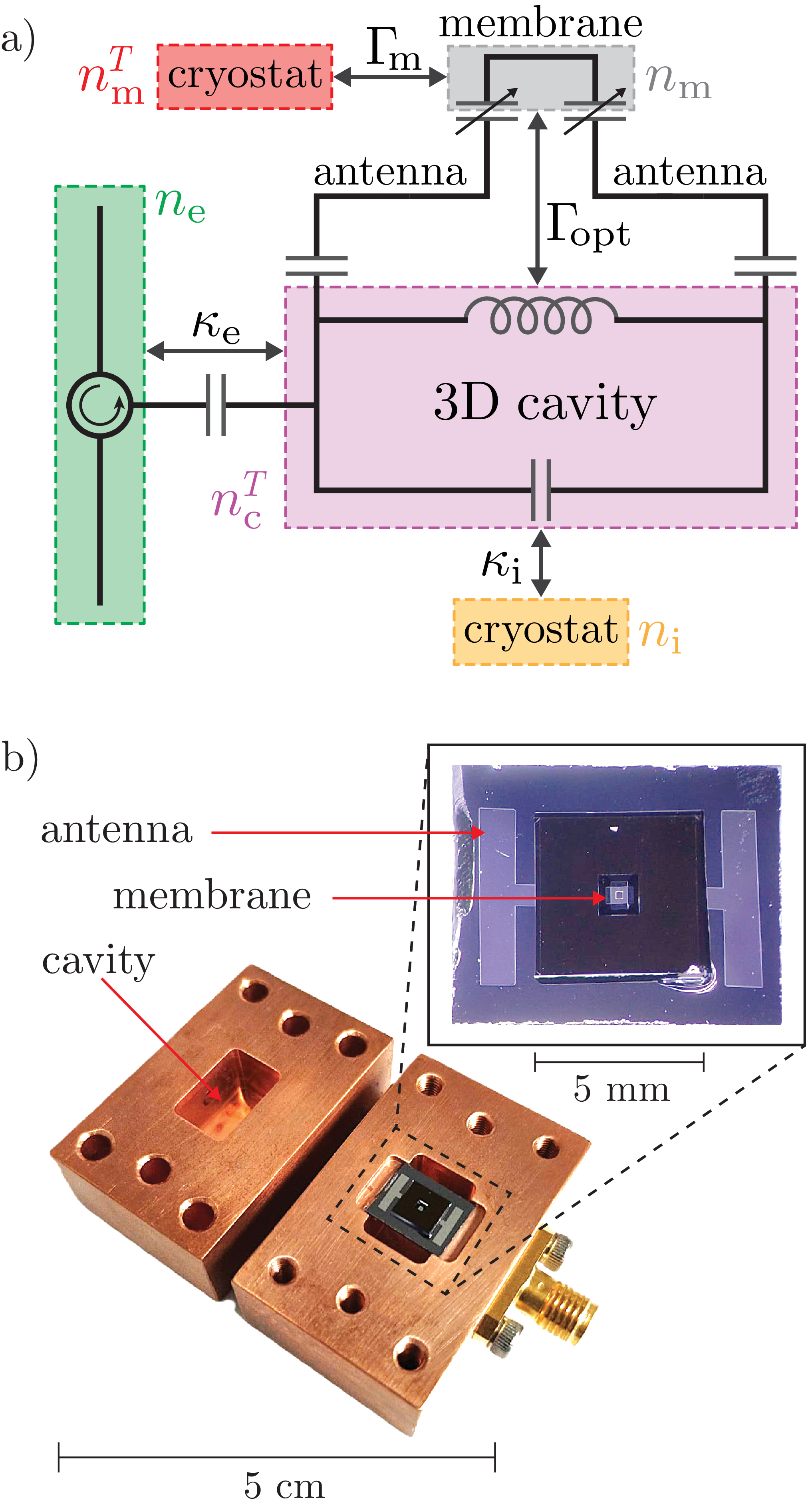}
    \caption{Optomechanical system: (a) Schematic of the equivalent circuit for the optomechanical system. The temperatures of relevant baths are shown in units of quanta with colored regions. Couplings between baths and their strengths are shown with labeled two-way arrows. (b)~ Photographs of a representative membrane and antenna device mounted to a 3D copper cavity.}
    \label{fig:CavSchematic}
\end{figure}

Our optomechanical system is a silicon nitride membrane coupled to a 3D microwave cavity\cite{liu2022quantum, rej2025near}, photographs of which are presented in Fig.~\ref{fig:CavSchematic}(b). The mechanical resonator is a metallized membrane which forms a capacitance with a nearby aluminum antenna, in turn relating the capacitance of the system to the membrane’s position. The chip assembly sits within a 3D cavity, and the cavity resonance frequency is modulated by the membrane’s oscillation, providing an optomechanical coupling with strength dependent on the membrane-antenna gap size.

The device is placed in a dry dilution refrigerator at a temperature of 10\,mK. Under these conditions, the system parameters are: cavity resonance frequency $\omega_{\mathrm{c}}/2\pi=~4.57$\,GHz, internal loss rate $\kappa_{\mathrm{i}}/2\pi~=~943$\,kHz, external coupling $\kappa_{\mathrm{e}}/2\pi~=~1.53$\,MHz, mechanical resonance frequency $\omega_{\mathrm{m}}/2\pi~=~707$\,kHz, and intrinsic mechanical damping $\Gamma_\mathrm{m}/2\pi~=~7$\,mHz.

\begin{figure*}[t]
    \centering
    \includegraphics[width=0.99\linewidth]{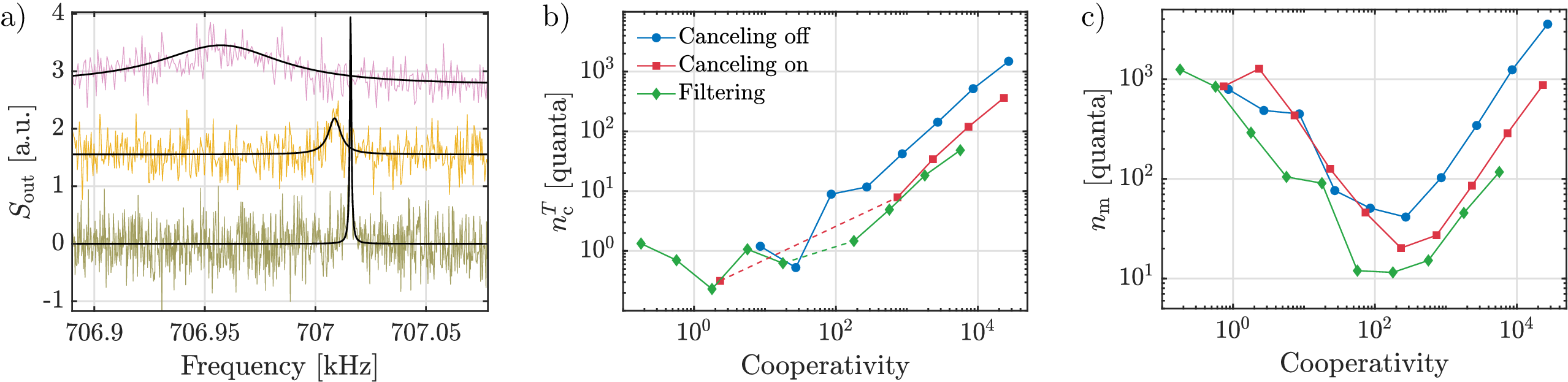}
    \caption{Extracting occupation numbers from the measured mechanical sidebands in the output spectrum: (a) Three example mechanical noise spectra for the filtered noise scenario with theoretical fits. From bottom to top the cooperativities are [18;~180;~1800]. Data are offset for clarity. (b) Fit parameters give the cavity bath occupations $n_\mathrm{c}^T$ and (c) oscillator occupations $n_\mathrm{m}$ as a function of cooperativity for the three noise scenarios. Missing data points occur when $n_\mathrm{c}^T = 0$, adjacent points to these are connected with dashed lines.}
    \label{fig:SBcool}
\end{figure*}

A single microwave tone is input into the cancellation circuit, and a signal analyzer is used to measure the noise properties of the output from the circuit in real-time, allowing for appropriate tuning of the FPGA parameters to ensure an optimized and symmetric cancellation profile. The tone is then sent to the cryostat, where the signal is reflected from the cavity via a circulator. For the purpose of sideband cooling, the pump frequency is set close to the red sideband ($\Delta \approx -\omega_\mathrm{m}$). At the room temperature output the signal undergoes heterodyne mixing with a local oscillator and the voltage time trace is recorded with a DAQ.

The starting value for $n_\mathrm{m}^T$ at low generator power, such that the optical damping is negligible, is determined by thermal calibration. In this instance, the sideband peak area is proportional to $n_\mathrm{m}^T$, which is measured for a range of cryostat temperatures. The regime in which peak area has a linear dependence with cryostat temperature indicates where the device and cryostat are in equilibrium, providing the temperature used to calculate $n_\mathrm{m}^T$. We estimate the device to equilibrate with the cryostat down to a temperature of 50\,mK when the pulse tube is turned of.

Similarly, the temperature gives the starting value for $n_\mathrm{c}^T$, which is vanishingly small for a microwave cavity at cryogenic temperatures. Background spectra are measured while the generator is off, the noise floors of which give the added noise quanta due to the measurement apparatus. The added noise is then subtracted from each spectrum measured in the experiment. When the noise floor of a spectrum measured with the pump on is equal to the background level, $n_\mathrm{c}^T = 0$. A measured increase in noise floor for certain pump powers is then attributed to an increase in the cavity bath temperature beyond its equilibrium value.

The procedure continues by measuring the anti-Stokes sideband for a range of generator powers. Lorentzian fits provide the total mechanical damping from each spectrum, such that the effective coupling can be determined using a linear fit to $\Gamma_\mathrm{opt}$ as a function of $P$, given by Eq.~(\ref{eq:Gammaopt}). The calibrated effective coupling for each generator power is then used as a set of parameters while simultaneously fitting the collection of spectra, where each spectrum has its own fit parameters for $n_\mathrm{i}$, $n_\mathrm{e}$ and $n_\mathrm{m}^T$.

\subsection{Sideband cooling results}

Three configurations of generator noise are compared during the experiment. The first is with the intrinsic generator noise and the second is with amplitude noise canceled at an offset of $\omega_\mathrm{m}$, henceforth referred to as the `canceling off' and `canceling on' scenarios respectively. The third method utilized room temperature cavity filters (referred to as the `filtering' scenario). Because the filters remove both amplitude noise and phase noise, they provide the best-case scenario for comparison. The use of filtering is currently the standard method used in most optomechanical experiments.

We measure the mechanical sidebands from the reflected signal, examples of which are shown in Fig.~\ref{fig:SBcool}(a), and extract the parameters $n_\mathrm{m}^T$, $n_\mathrm{i}$ and $n_\mathrm{e}$ from theoretical fits, which are then used to calculate $n_\mathrm{c}^T$ using Eq.~(\ref{eq:ncT}) and $n_\mathrm{m}$ using Eq.~(\ref{eq:nm}).

Our main focus is on how the canceling or filtering affects the external contribution, $n_\mathrm{e}$, to the cavity bath temperature. Fig.~\ref{fig:SBcool}(b) shows the cavity bath temperatures, $n_\mathrm{c}^T$, for each scenario. Above a cooperativity of 100, $n_\mathrm{c}^T$ increases linearly in all scenarios with varying gradients. The intrinsic generator noise causes $n_\mathrm{c}^T$ to rise approximately 3.5 times faster with cooperativity than in the canceling on scenario. Filtering the noise reduces the rate of heating by an additional factor of 2.

The effect of external heating is further demonstrated by observing the oscillator population $n_\mathrm{m}$ in the three scenarios as a function of cooperativity, see Fig.~\ref{fig:SBcool}(c). At low cooperativity, the oscillator is in equilibrium with the phonon bath, with $n_\mathrm{m}$ decreasing as the cooperativity becomes larger, before increasing again as the cavity noise becomes significant. The minimum occupation numbers we measure are [11.5;~20.2;~41.4] for the filtering, canceling on, and canceling off scenarios respectively, demonstrating that the cancellation technique is advantageous for cooling as compared to not employing any noise reduction measures, reducing the minimum $n_\mathrm{m}$ by more than a factor of 2. The $n_\mathrm{m}$ values at the largest cooperativities are also lower with the cancellation than they are without it.

\section{Conclusion}

In conclusion, we have demonstrated an accessible and simple cancellation technique to reduce the amplitude noise of a microwave tone. The method allows for tunable frequency and bandwidth, and we measured a 13\,dB reduction in total noise at a 2\,MHz offset from a 4\,GHz tone, with cancellation down to the phase noise level. Furthermore, we investigated the effect of noise cancellation in a sideband cooling experiment. We demonstrated that the canceling reduces external cavity noise and allows for a lowering of the minimum phonon occupation of a mechanical oscillator.

Beyond sideband cooling, this technique has applications for any experiments which use large microwave powers and where the associated generator noise is of a concern. The technique is favorable as compared to the use of cavity filters when easily tunable cancellation frequency and bandwidths are required, especially when using bandwidths smaller than a few hundred kilohertz that are typical of room temperature cavity filters. Finally, as the phase noise performance of microwave sources improves over time, techniques to reduce amplitude noise are becoming more relevant for technological advancements in low noise microwave generation.


\begin{acknowledgments}
We acknowledge the facilities and technical support of Otaniemi research infrastructure for Micro and Nanotechnologies (OtaNano). This work was supported by the Research Council of Finland (contract 352189), and by the European Research Council (contract 101019712). The work was performed as part of the Research Council of Finland Centre of Excellence program (contracts 352932, and 336810). We acknowledge funding from the European Union’s Horizon 2020 research and innovation program under grant agreement 824109, the European Microkelvin Platform (EMP), and the QuantERA II Programme (contract 13352189). This work has received funding from the European Union’s Research and Innovation Programme, Horizon Europe, under the Marie Skłodowska-Curie Grant Agreement No. 101198933 (mGramm). Y. L. acknowledges funding by the Beijing Municipal Science and Technology Commission (grant Z221100002722011).
\end{acknowledgments}

\section*{Author Declarations}

\subsection*{Conflict of Interest}

The authors have no conflicts to disclose.

\subsection*{Author Contributions}

\textbf{Joe Depellette:} Conceptualization (equal), data curation (lead), formal analysis (lead), investigation (lead), software (supporting), validation (equal), visualization (lead), writing - original draft (lead), writing – review \& editing (lead). \textbf{Ewa Rej:} Formal analysis (supporting), investigation (supporting), visualization (supporting), writing – review \& editing (supporting). \textbf{Matthew Herbst:} Formal analysis (supporting), writing – review \& editing (supporting). \textbf{Richa Cutting:} Investigation (supporting). \textbf{Yulong Liu:} Resources (supporting). \textbf{Mika A. Sillanpää:} Conceptualization (equal), data curation (supporting), formal analysis (supporting), funding acquisition (lead), project administration (lead), resources (lead), software (lead), validation (equal), writing – review \& editing (supporting).

\section*{Data Availability}

The data that support the findings of this study are available from the corresponding author upon reasonable request.

\section*{References}

\bibliography{References}

\clearpage
\pagebreak



\end{document}